# On the Calculation of Lorenz Numbers for Complex Thermoelectric Materials

Xufeng Wang[1], Vahid Askarpour[2], Jesse Maassen[2], and Mark Lundstrom[1]

[1]Purdue University, West Lafayette, IN USA

[2]Dalhousie University, Halifax, NS Canada

**Abstract-** A first-principles informed approach to the calculation of Lorenz numbers for complex thermoelectric materials is presented and discussed. Example calculations illustrate the importance of using accurate band structures and energy-dependent scattering times. Results obtained by assuming that the scattering rate follows the density-of-states show that in the non-degenerate limit, Lorenz numbers below the commonly assumed lower limit of $2(k_B/q)^2$ can occur. The physical cause of low Lorenz numbers is explained by the shape of the transport distribution. The numerical and physical issues that need to be addressed in order to produce accurate calculations of the Lorenz number are identified. The results of this study provide a general method that should contribute to the interpretation of measurements of total thermal conductivity and to the search for materials with low Lorenz numbers, which may provide improved thermoelectric figures of merit, $zT$.

## 1. Introduction

Knowledge of the Lorenz number, $L$, is essential for interpreting measurements of total thermal conductivity, and it has been recently noted that identifying materials with low Lorenz numbers may provide a path to higher thermoelectric figures of merit [1]. Measurements of $L$ can be done [2-5], but because such measurements are involved, they are not routinely performed. It has been recently pointed out that for parabolic energy bands with acoustic deformation potential (ADP) scattering, there is a direct relation between $L$ and the routinely measured Seebeck coefficient, $S$ [6]. The $L$ vs. $S$ characteristic is independent of the value of the effective mass and independent of temperature when bipolar conduction is negligible. This approach provides a convenient way to determine $L$ from a measured $S$, but the assumption of a simple band structure and scattering process raises concerns about its generality. Calculations of $L$ can be done [7]), but they typically assume simplified band structures and scattering processes. Numerical solutions to the Boltzmann Transport Equation in the Relaxation Time

Approximation [8, 9] should provide the most accurate predictions of $L$ when informed by rigorous treatments of electron scattering [8, 10-14]. In this paper, we present a first-principles informed approach for calculating $L$ for complex thermoelectric materials. As an example of the approach, we present calculations for $Bi_2Te_3$ that assume the scattering rate follows the density-of-states (DOS). These calculations show that for non-degenerate conditions, $L$ can be lower than the often-assumed lower limit of $2(k_B/q)^2$, and they contribute to the understanding of the physical mechanisms behind low Lorenz numbers. Under degenerate conditions, $L$ can deviate from the Sommerfeld limit of $(\pi^2/3)(k_B/q)^2$, which often describes metals and heavily doped semiconductors [15]. The addition of bipolar conduction further complicates the calculation and can result in unusually large Lorenz numbers. Finally, we discuss uncertainties in the calculation of $L$ in order to guide future work directed at the accurate calculation of $L$ for complex thermoelectric materials.

## 2. Approach

Our approach is to calculate the $L$ vs. $S$ characteristic for complex thermoelectric materials and to compare the results to analytical calculations for simple, parabolic energy bands. It is convenient to use the $L$ vs. $S$ characteristic because both $L$ and $S$ depend on the energy dependence of the scattering time, but not on its magnitude. The standard expressions for the thermoelectric transport coefficients that result from a relaxation time approximation solution to the Boltzmann transport equation are:

$$\sigma = \int_{-\infty}^{+\infty} \sigma'(E) dE \tag{1a}$$

$$S = -\frac{1}{qT} \frac{\int_{-\infty}^{+\infty} (E-E_F)\sigma'(E) dE}{\int_{-\infty}^{+\infty} \sigma'(E) dE} \tag{1b}$$

$$\kappa_0 = \frac{1}{q^2 T} \int_{-\infty}^{+\infty} (E-E_F)^2 \sigma'(E) dE \tag{1c}$$

$$\kappa_e = \kappa_0 - T\sigma S^2 \tag{1d}$$

$$L \equiv \frac{\kappa_e}{\sigma T}, \tag{1e}$$

where $S$ is the Seebeck coefficient, $\kappa_o$ is the short-circuit electronic thermal conductivity, $\kappa_e$ is the open-circuit electronic thermal conductivity, and the differential conductivity is

$$\sigma'(E) = q^2 \Xi(E)\left(-\frac{\partial f_0}{\partial E}\right), \qquad (1f)$$

and the diagonal component of the transport function is [9]

$$\Xi_{xx}(E) \equiv \sum_{\vec{k}} v_x^2 \tau_m(E) \delta(E - E_k), \qquad (1g)$$

where an energy-dependent scattering time has been assumed. As shown by Mahan and Soho, all thermoelectric transport coefficients are determined by the transport function [10], so the Lorenz number, $L$, which is a ratio of transport coefficients as given by (1e), is also determined by the transport function.

We will take (1e) as the definition of the Lorenz number, which is different from the Wiedemann-Franz Law, which assigns a specific value, $L = L_0 = (\pi^2/3)(k_B/q)^2$ to the Lorenz number [17]. It is common, however, to refer to $\kappa_e/\sigma = LT$ as the "Wiedemann-Franz Law" and apply it to semiconductors for which $L$ is rarely $L_0$ as well as to metals (see, for example, [18]). As Mahan and Bartkowiak state, the Wiedemann-Franz Law should be regarded as a rule of thumb [18]. Violations occur, especially for electronic structures and scattering processes that produce energetically sharp transport functions. One example is the one-dimensional organic crystals considered in [2]. Our goal in this paper is to present a method to compute the Lorenz number as defined by (1e) for thermoelectric materials using the best available information about band structures and scattering times.

In this work, we solve eqns. (1) but find it convenient to write the transport function in Landauer form as [16] (see also the Supplementary Information)

$$\Xi_{xx}(E) = \frac{2}{h}(M(E)/A)\lambda(E), \qquad (2)$$

where $M(E)/A$ is the number of conduction channels per unit cross-sectional area vs. energy. We compute $M(E)/A$ from a DFT-generated band structure using the open source tool, LanTraP 2.0 [17]. The energy-dependent mean-free-path for backscattering is also needed; it is defined as [16]

$$\lambda(E) \equiv 2\frac{\langle v_x^2(E)\rangle}{\langle |v_x(E)|\rangle}\tau_m(E), \tag{3}$$

where $\tau_m(E)$ is the momentum relaxation time, and the quantity, $\langle v_x^2\rangle/\langle |v_x|\rangle$, is an angle-averaged velocity and is computed as a function of energy from the DFT-generated band structure. For acoustic deformation potential (ADP) scattering in the elastic limit, the scattering rate is isotropic, equal to the momentum relaxation rate, and proportional to the density of states:

$$\frac{1}{\tau(E)} = \frac{1}{\tau_m(E)} = K_{el-ph}\,DOS(E). \tag{4}$$

More generally, it is often found that the electron-phonon scattering rate follows the DOS in non-polar semiconductors [21, 9] and even in polar semiconductors when the electron-phonon coupling is strongly screened (see Sec. 4 Discussion). The density-of-states can be computed directly from the numerical band structure. The electron-phonon coupling parameter, $K_{el-ph}$, is proportional to the deformation potential squared, but its value does not need to be specified because it appears in both the numerator and denominator of the expressions and, therefore, drops out when computing both $L$ and $S$ (which is one reason for focusing on the $L$ vs. $S$ characteristic). When there are multiple scattering mechanisms, the relative strength of each mechanism must be known.

To illustrate the principles involved, we will begin in Sec. 3 with a parabolic conduction band for which eqns. (1) can be evaluated analytically when power law scattering for the energy dependent mean-free-path for backscattering is assumed,

$$\lambda(E) = \lambda_0 (E-E_C)^r / k_B T. \tag{5}$$

In this expression, $r$ is a characteristic exponent determined by the electron scattering processes. The numerical calculations that will be presented in Sec. 3 evaluate $\lambda(E)$ directly from (3) and (4) and do not assume the power law form of (5).

Ignoring bipolar conduction and assuming parabolic energy bands, eqns. (1) can be evaluated to find [18]

$$S = -\left(\frac{k_B}{q}\right)\left(\frac{(r+2)\mathcal{F}_{r+1}(\eta_F)}{\mathcal{F}_r(\eta_F)} - \eta_F\right) = S'(k_B/q) \tag{6a}$$

$$L = \left(\frac{k_B}{q}\right)^2 \frac{\Gamma(r+3)}{\Gamma(r+2)} \left[(r+3)\frac{\mathcal{F}_{r+2}(\eta_F)}{\mathcal{F}_r(\eta_F)} - (r+2)\left(\frac{\mathcal{F}_{r+1}(\eta_F)}{\mathcal{F}_r(\eta_F)}\right)^2\right] = L'(k_B/q)^2 \quad (6b)$$

where

$$\eta_F = (E_F - E_C)/k_B T \quad (7)$$

is the reduced Fermi level. In (6a) and (6b), we defined the dimensionless Seebeck coefficient and Lorenz number, $S'$ and $L'$ respectively. In these expressions,

$$\mathcal{F}_j(\eta_F) = \frac{1}{\Gamma(j+1)} \int_0^\infty \frac{\eta^j d\eta}{1 + e^{\eta-\eta_F}} \quad (8)$$

is the Fermi-Dirac integral of order $j$ as defined by Blakemore [19]. Note that the value of the effective mass and the magnitude of the mean-free-path, $\lambda_0$, do not appear in these expressions. The $L$ vs. $S$ characteristic computed from eqns. (6) depends only on the scattering exponent, $r$, and is independent of temperature when bipolar conduction is negligible.

## 3. Results

In this section, we compute the $L$ vs. $S$ characteristic in three different ways. The first is an analytical calculation that assumes parabolic bands and power law scattering. The second is a full, numerical treatment of silicon (where the bands are approximately parabolic), and the third is a numerical treatment of Bi$_2$Te$_3$, which has a complex band structure. Comparing these three cases sheds light on how the transport distribution determines $L$.

*Analytical treatment (parabolic bands and power law scattering)*

Figure 1 shows the computed $L$ vs. $S$ characteristic for various values of $r$. Acoustic deformation potential (ADP) scattering (commonly thought to be the dominant scattering mechanism in thermoelectric materials [24]) corresponds to $r = 0$. For degenerate semiconductors, $\eta_F \gg 0$, $S \to 0$. In the degenerate limit, Fig. 1 shows that the normalized Lorenz numbers, $L' = L/(k_B/q)^2$, approach the Sommerfeld limit of $\pi^2/3$ [15]. For non-degenerate semiconductors, $|S|$ is large, and Fig. 1 shows that for $r = 0$, $L'$ approaches the expected limit of 2. For $r = 2$, which corresponds to ionized impurity scattering, $L' \to 4$ in the non-degenerate limit. For $r < 0$, Fig. 1 shows that $L' < 2$ in the non-degenerate limit.

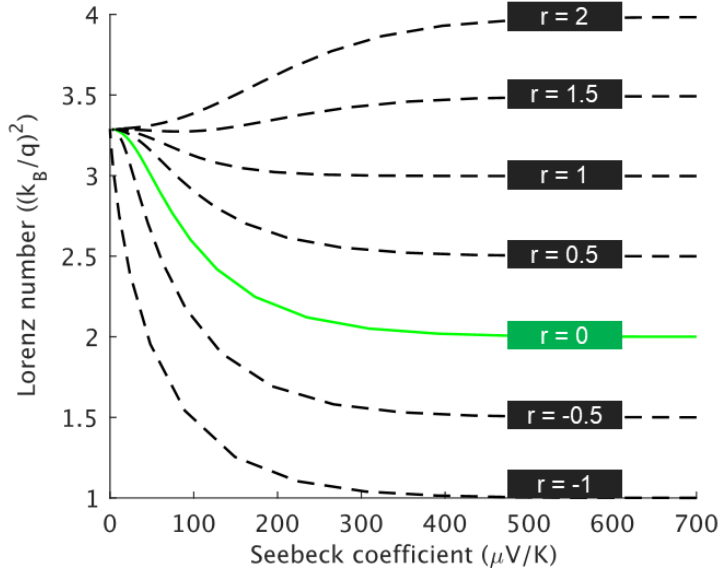

Fig. 1 The $L$ vs. $S$ characteristic for parabolic energy bands showing several different values of the power law scattering exponent, $r$. For $r = 0$, which corresponds to ADP scattering in 3D, the results are identical to those in Ref [6].

As given by eqns. (1), the thermoelectric transport coefficients are integrals over energy of various powers of $(E - E_F)$. The weighting factor is the transport function, $\Xi(E)$, times the Fermi window, $(-\partial f_0/\partial E)$. In a Landauer picture, it can be shown that the transport distribution is proportional to the number of channels for conduction, $M(E)$, times the mean-free-path for backscattering, $\lambda(E)$ [16]. For three-dimensional electrons in parabolic bands, one can show that $M(E)$ increases linearly with energy, so for power law scattering,

$$\Xi(E) \propto M(E)\lambda(E) \propto (E - E_C)^{r+1} . \qquad (9)$$

When $r = 0$ (ADP scattering in parabolic bands), the transport distribution increases linearly with energy and $L' = 2$ results in the non-degenerate limit. For $r < 0$, the transport distribution increases sub-linearly with energy, and $L' < 2$ results in the non-degenerate limit.

To understand why the Lorenz number is sensitive to the characteristic exponent, $r$, it is useful to write the transport parameters in terms of averages over energy of moments of $(E - E_F)$. For $S'$ and $L'$ the result is [18]

$$S' = -\left\langle \left( \frac{E-E_F}{k_B T} \right) \right\rangle \tag{10a}$$

$$L' = \left\langle \left( \frac{E-E_F}{k_B T} \right)^2 \right\rangle - \left\langle \left( \frac{E-E_F}{k_B T} \right) \right\rangle^2, \tag{10b}$$

where the brackets denote an average over energy. Equations (10) show that the Lorenz number emphasizes the higher energies more that the Seebeck coefficient does. For $r > 0$, higher energies are weighted more than for $r = 0$, so $L' > 2$. For $r < 0$, higher energies are weighted less than for $r = 0$, so $L' < 2$. When we turn next to the numerical computation of the $L$ vs. $S$ characteristic, we will see that the non-degenerate limits can be explained by examining the shape of the numerically calculated transport distribution.

As a final note, we point out that eqns. (1) assume three-dimensional electrons, but if we use the corresponding equations for 2D and 1D electrons [18], we find that for ADP scattering in parabolic bands, the $L$ vs. $S$ characteristic is identical in 1D, 2D, and 3D. This can be easily seen from (2). The mean-free-path is proportional to velocity times scattering time, and the scattering time is inversely proportional to the DOS. The number of channels is proportional to velocity times the DOS [19, 22], so for parabolic bands, we find

$$\Xi_{xx}(E) \propto \left( M(E)/A \right) \times \lambda(E) \propto \upsilon(E) D(E) \times \upsilon(E)/D(E) \propto \upsilon^2(E) \propto (E-E_C). \tag{11}$$

Since the transport function is independent of dimension for ADP scattering in parabolic bands, all thermoelectric transport coefficients are independent of dimension.

*Numerical treatment (Si with full, numerical energy bands)*
In this case, we evaluate eqns. (1) assuming a DFT-generated band structure. Figure 2 shows the density-of-states computed from numerical band structure. The relevant portion where the Fermi window overlaps the density-of-states, lies very close to the Fermi level, which is chosen to be at the valence band edge in this case. The corresponding distribution of channels, $M(E)$, energy-dependent mean-free-path, $\lambda(E)$, and transport distribution, $\Xi(E)$, are also shown in Fig. 2.

The density-of-states shown in Fig. 2a confirms that the conduction and valence bands are nearly parabolic in Si. The onset of a second conduction band about 0.2 eV above the bottom of the conduction band can also be seen. The distribution of channels, $M(E)$, shown in Fig. 2b is approximately linear with energy, as expected for parabolic energy bands [16, 18]. A change in slope for $M(E)$ is seen in the conduction band where the second conduction band begins. The valence band for Si is known to have a complex, warped shape [20], but it has also been shown that it can be described by an equivalent parabolic band [21], which is confirmed by the linear behavior of $M(E)$ in the valence band. Figure 2c shows the energy-dependent mean-free-path numerically evaluated according to eqns. (3) and (4). Note that for a parabolic energy band with ADP scattering, $\lambda(E) = \lambda_0$ is independent of energy. The numerically calculated $\lambda(E)$ is approximately independent of energy. Numerically evaluating $\lambda(E)$ presents challenges near the band edge because $\lambda(E) \sim \upsilon(E)\tau_m(E)$, and the velocity approaches zero as the scattering time becomes infinite. As discussed in the Supplementary Information, the use of a fine grid near the band edge is essential. Finally, Fig. 2d shows the numerically computed transport distribution, $\Xi(E)$, according to eqn. (2). The use of a fine energy grid near the band edge produces a smooth $\Xi(E)$, which is then used in (1) to evaluate the thermoelectric transport coefficients.

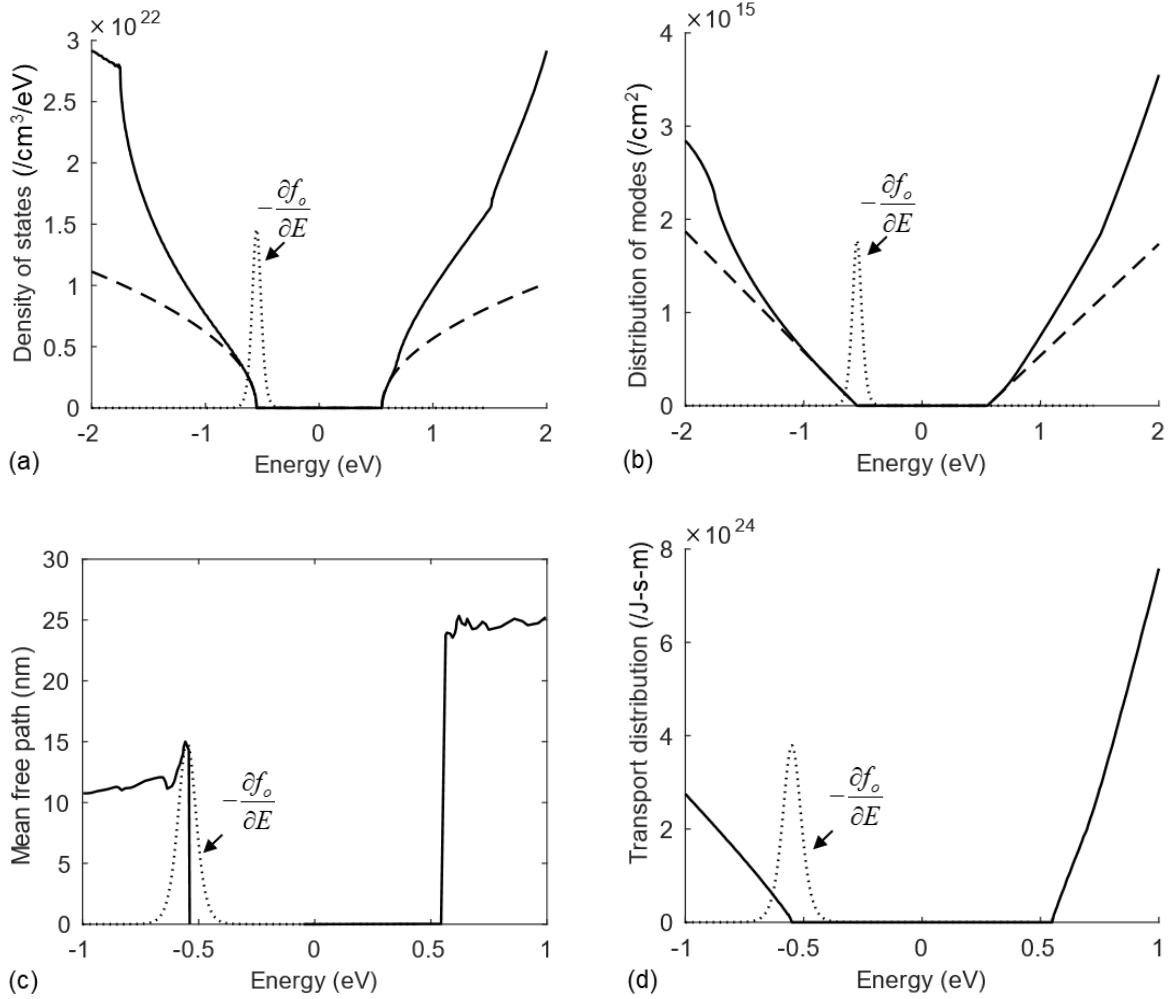

Fig. 2 Key quantities computed from the numerical band structure of Si. 2a): The density-of-states (solid line) and its parabolic band fit (dash line). Fig. 2b): The distribution of channels (solid line), $M(E)$, and its parabolic band fit (dashed line). Fig. 2c): the energy-dependent mean-free-path, $\lambda(E)$, and Fig. 2d): The transport distribution, $\Xi(E)$. The Fermi window, which is shown as a dotted line centered at the Fermi level indicates the relevant range of energies. The Fermi level is set at the valence band edge in these plots. All quantities were extracted from a DFT-generated band structure using Quantum Espresso [22] (see Supplementary Information). The bandgap is 1.1 eV with midgap at $E = 0$ eV.

The $L$ vs. $S$ characteristic for Si is computed by using the $M(E)$ and $\lambda(E)$ presented in Fig. 2 and calculated from eqns. (1). Three cases are shown in Fig. 3: i) the conduction band (symbols), ii) the valence band (solid line), and iii) the analytically calculated parabolic band reference (dashed line). The numerical results for both the conduction and valence bands of Si are very close to those obtained from the analytical expressions. This might be expected for the conduction band, which consists of six parabolic ellipsoids, but it is not so apparent why the complex valence band behaves like a simple, parabolic band. The reason is provided by Mecholsky et al., who show that the warped valence band is mathematically equivalent to a set of ellipsoidal bands [21]. The observation from Fig. 2d that the transport distribution is nearly linear with energy over most of the Fermi window, indicates that a parabolic band is a good approximation. The numerical calculations were done at 300 K, but the results are only weakly dependent on temperature (see Supplementary Information). (For the analytical model, the $L$ vs. $S$ characteristic is independent of temperature.) Note that bipolar effects are not considered here, but are later discussed in Sec. 4.

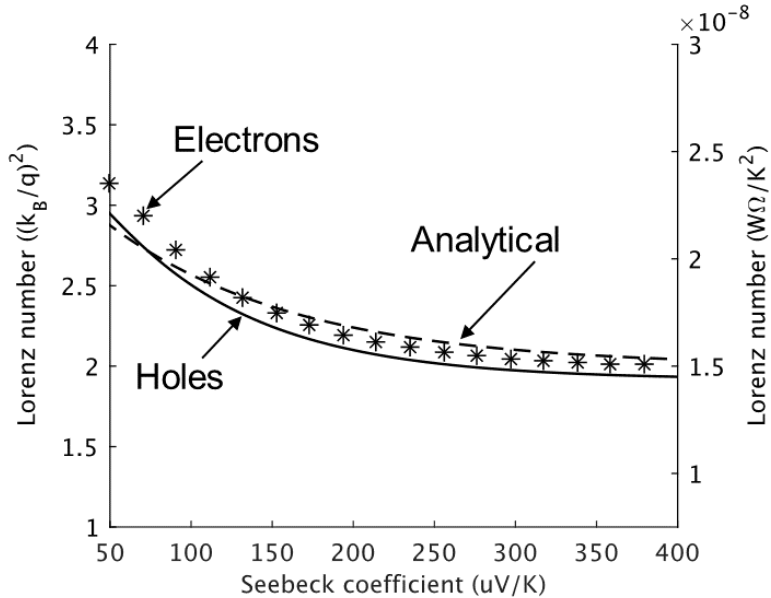

Fig. 3 The $L$ vs. $S$ characteristic computed numerically for Si. Symbols: the conduction band with DOS scattering. Solid line: the valence band with DOS scattering. Dashed line: the analytically calculated parabolic band reference.

*Numerical treatment (Bi$_2$Te$_3$ with full, numerical energy bands)*

In this case, we evaluate eqns. (1) assuming a DFT-generated band structure for Bi$_2$Te$_3$. Figure 4 shows the density-of-states computed from numerical band structure along with the corresponding distribution of channels, $M(E)$, energy-dependent mean-free-path, $\lambda(E)$, and transport function, $\Xi(E)$.

The non-parabolic energy bands are evident in the density-of-states shown in Fig. 4a. Nevertheless, the distribution of channels, $M(E)$, shown in Fig. 4b is roughly linear with energy over the Fermi window. As shown in Fig. 4c, the drop in the magnitude of $\lambda(E)$ for energies away from the band edge is much stronger for the valence band than for the conduction band. Finally, Fig. 4d shows the numerically computed transport distribution, $\Xi(E)$, which is proportional to the product of $M(E)$ and $\lambda(E)$. This figure shows that for the conduction band, the increase in $\Xi(E)$ above the conduction band minimum is slightly sublinear. For the valence band, however, the increase in $\Xi(E)$ for energies below $E_V$ is distinctly sublinear. The analytical results presented earlier showed that a sub-linear increase in $\Xi(E)$ with energy leads to $L' < 2$ for non-degenerate conditions. Based on the results shown in Fig. 4d, we might expect $L'$ for the conduction band to be slightly less than 2 for non-degenerate conditions, but $L'$ for the valence band is expected to be distinctly less than 2 for non-degenerate conditions.

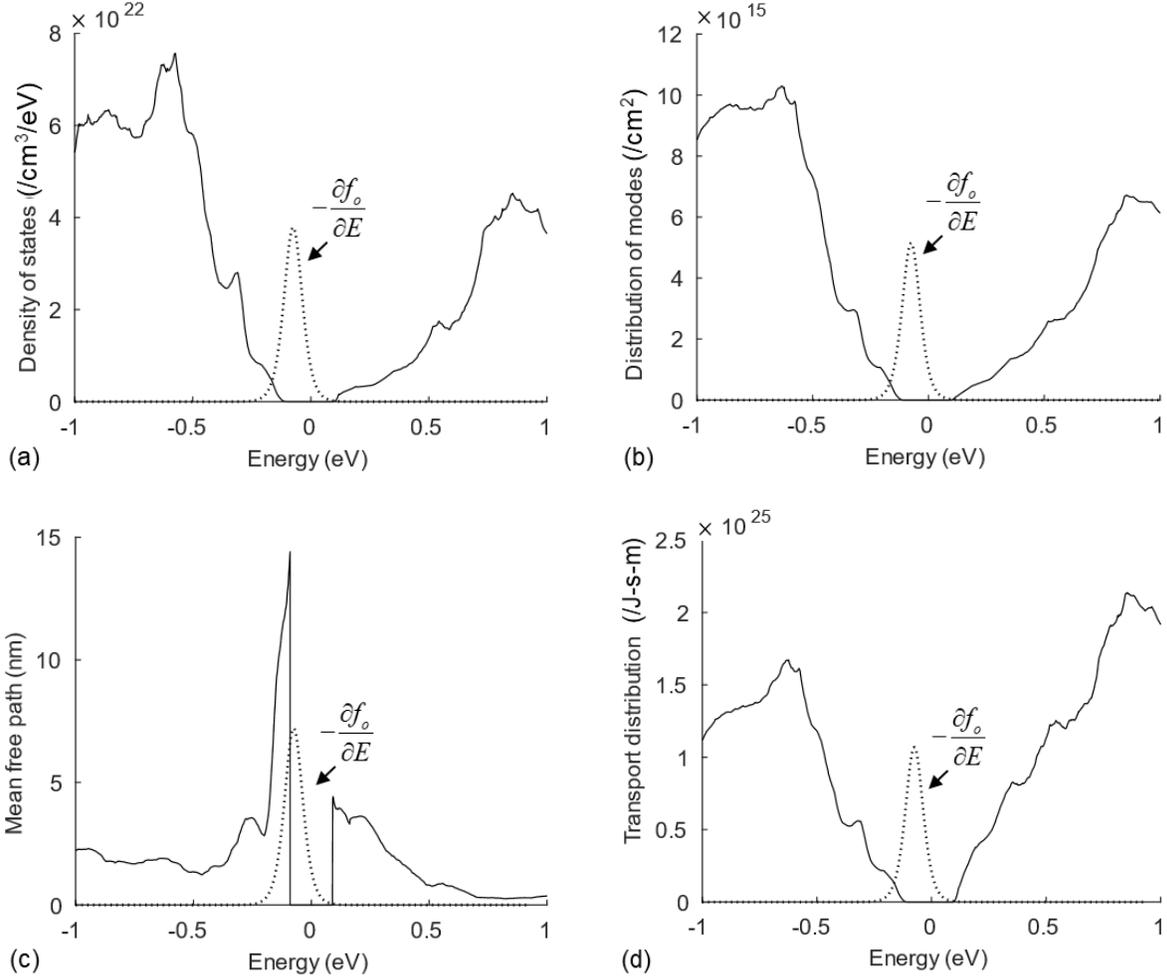

Fig. 4 Key quantities computed from the numerical band structure of Bi$_2$Te$_3$. Fig. 4a): The density-of-states. Fig. 4b): The distribution of channels, $M(E)$. Fig. 4c): the energy-dependent mean-free-path, $\lambda(E)$. Fig. 4d): The transport distribution, $\Xi(E)$. The Fermi window is shown as a dashed line centered at the Fermi level and indicates the relevant range of energies. The Fermi level is set at the valence band edge in these plots. All quantities were extracted from a DFT-generated band structure using Quantum Espresso (see Supplementary Information). The bandgap is 0.18 eV with midgap at $E = 0$ eV.

The *L* vs. *S* characteristic for Bi$_2$Te$_3$ as computed by using the $\Xi(E)$ shown in Fig. 4 with eqns. (1) is presented in Fig. 5. Three cases are shown: i) the conduction band (symbols), ii) the valence band (solid line), and iii) the analytically calculated parabolic band reference (dashed line). As expected, the numerical result for the conduction band is somewhat less than the parabolic band result. For the valence band, however, $L'$ is distinctly less than 2 for non-degenerate conditions. As discussed above, the behavior observed in Fig. 5 can be explained in terms of the transport function.

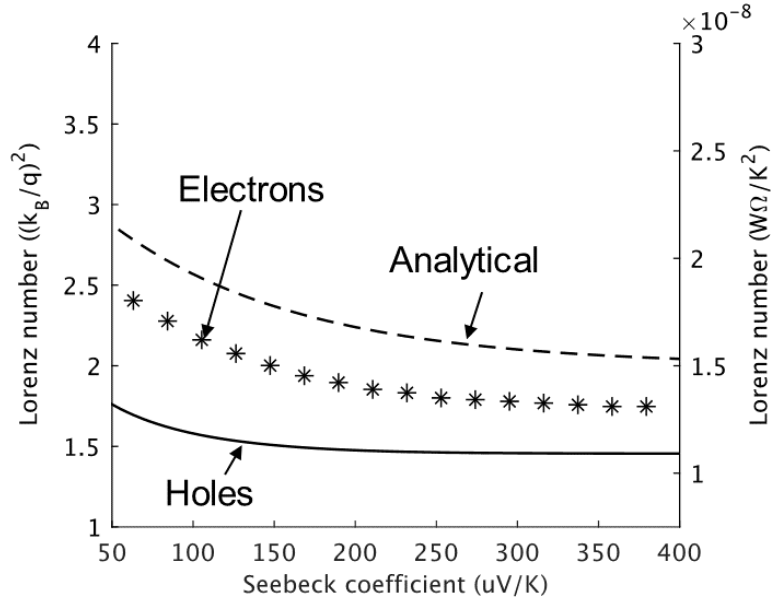

Fig. 5 The $L$ vs. $S$ characteristic computed numerically for $Bi_2Te_3$. Symbols: the conduction band with DOS scattering. Solid line: the valence band with DOS scattering. Dashed line: the analytically calculated parabolic band reference.

## 4. Discussion

Four issues will be discussed in this section. The first is the sensitivity of the results to the band structure. The second issue is the sensitivity of the results to scattering processes, and the third issue concerns the influence of bipolar effects, which have been neglected in the analyses discussed above. Finally, numerical issues are briefly discussed here and at more length in the Supplementary Information.

### *Effects of band structure and scattering*

The differences between the analytically and numerically calculated $L$ vs. $S$ characteristics are the result of both band structure and scattering. In this section, band structure and scattering effects are examined for Si and $Bi_2Te_3$. Figure 6a shows the $L$ vs. $S$ characteristic for hole-only conduction in the valence bands of Si and compares different treatments of scattering with the analytical results that assume parabolic bands and a constant MFP. When the MFP is computed by assuming that the scattering rate follows the density of states, the small difference between the analytical result and the numerical result is mainly due to the carrier velocity averages as given by (3). The use of a Constant MFP approximation (CMFPA) also produces results that are close to the analytical results. The warped valence bands produce a linear $M(E)$ but the energy-dependent angle-averaged velocity differs from that of a parabolic band. It may be surprising that the numerical results agree so closely with the analytical results because the valence bands of Si are complicated, but as discussed by Mecholsky et al., the warped valence bands are mathematically equivalent to a set of ellipsoidal bands [21]. Finally, we consider a commonly used approximation when calculating thermoelectric properties – the constant relaxation time approximation (CRTA). This assumption simplifies the analysis but is non-physical. Figure 6a shows that the results for the CRTA are much different from the others. The use of a constant relaxation time is non-physical - even for parabolic bands.

Next, we turn to Fig. 6b, the case of $Bi_2Te_3$, which has a band structure that is more complicated and significantly non-parabolic as compared to silicon. In this case, the results for a numerical band structure with a constant MFP or a constant scattering time are close to the analytical, parabolic band model. Significantly lower Lorenz numbers are, however, obtained when DOS scattering is assumed. DOS scattering is thought to be the most physical of the three scattering models assumed.

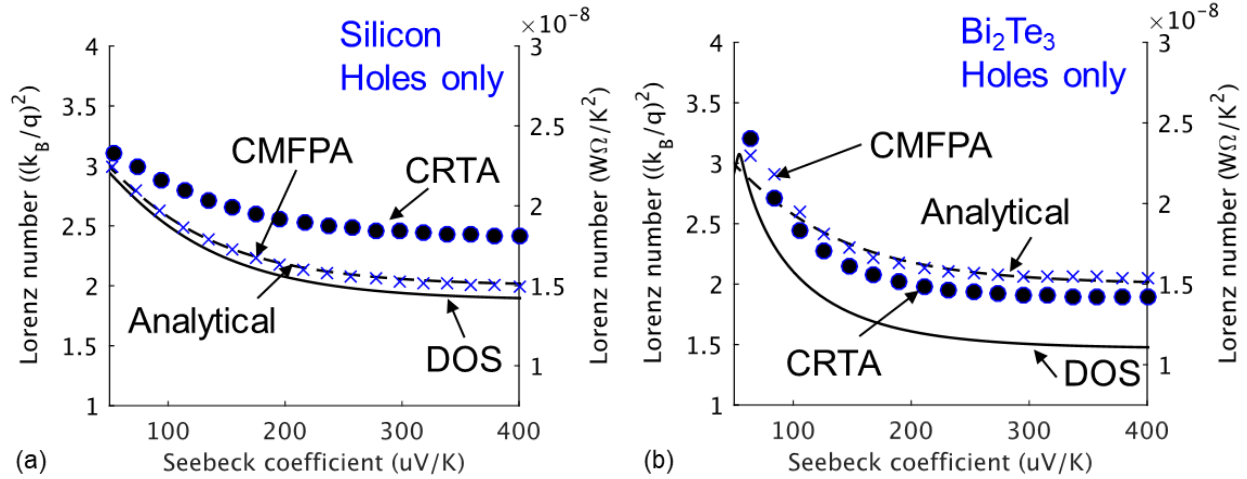

Fig. 6 The *L* vs. *S* characteristic for (a) silicon and (b) $Bi_2Te_3$ with constant mean free path approximation ("CMFPA", crosses), constant relaxation time approximation ("CRTA", solid circles), and scattering rate that is proportional to the DOS ("DOS", solid). In both cases, a reference calculation with single parabolic band and constant mean free path ("Analytical", dashed) is shown. These calculations treat the valence band alone and do not consider bipolar conduction.

*DOS scattering*

Although the technique described in this paper can make use of arbitrary numerical tables of band structure and energy-dependent scattering times, we have illustrated it by assuming that the scattering rate follows the DOS. This has been found to be a good approximation for non-polar semiconductors (e.g. [21] and [9]), but its validity for complex thermoelectric materials is less clear. Techniques to rigorously evaluate electron-phonon scattering rates in complex materials are available [11-14] and can address this question. Calculations for p-type SnSe are shown in Fig. 7 below. (For a description of the methods used, see Supplementary Information.) Each point in the plot represents the scattering rate for a particular *k*-state with all relevant acoustic and optical phonon modes included. The polar phonon interactions are screened according to the Debye length with the Fermi level being near the valence band edge. The results show that the scattering rate is *k*-dependent with the variation with *k* increasing as the energy increases. A comparison of the scattering rate with the momentum relaxation rate shows that the two are approximately the same. Without screened polar interactions, the scattering rate is significantly larger than the momentum relaxation rate. The solid line is the computed DOS for SnSe. These results show that even for a polar material (with relatively high carrier concentration; $4\times10^{19}$ cm$^{-3}$ in this case) with a complex band structure, the assumption that the momentum relaxation time follows the DOS is reasonable and much better than the often-assumed constant scattering time. In general, however, a detailed examination of material-specific scattering processes will be needed to provide accurate scattering times for the approach described in this paper.

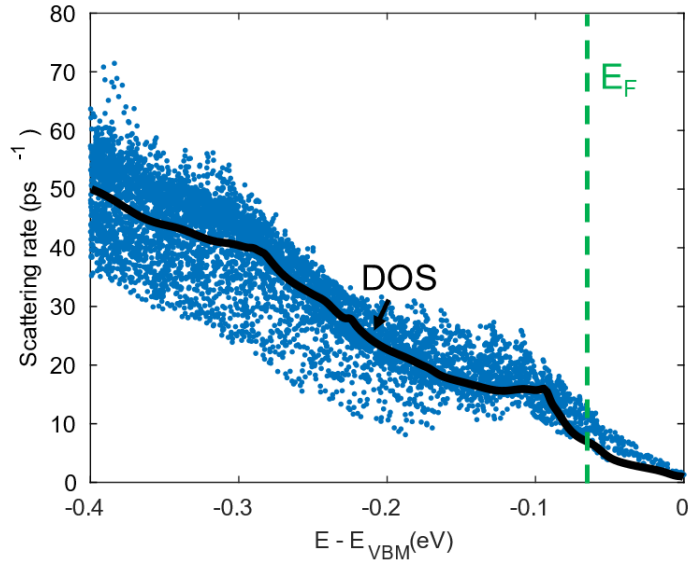

Fig. 7 Computed electron-phonon scattering rate for p-type SnSe. (See Supplementary Information for a description of the techniques used.) Each point represents the momentum scattering rate for a particular k-state. The solid line is the computed DOS for SnSe. The Fermi level is 64.6 meV below the top of the valence band, and corresponds to a hole concentration of $4\times10^{19}$ cm$^{-3}$.

*Lorenz number with bipolar conduction*

The study so far has considered *L* vs. *S* when only one type of carrier is present. This is acceptable at 300 K for Si, which has a bandgap of 1.1 eV, but for many common TE materials bipolar effects can become important at the temperatures of interest (e.g. Bi$_2$Te$_3$ at room temperature with $E_G \approx 0.1-0.2 \text{ eV}$). Computing an *L* vs. *S* characteristic in the presence of bipolar conduction adds uncertainties because in addition to knowledge of the majority carrier transport properties and bandgap, minority carrier properties are also needed (e.g. the MFP and effective mass in a parabolic band model or the minority carrier band structure and scattering rate in a full, numerical model). We can illustrate the considerations by looking at $Bi_2Te_3$ for which considerable experimental data is available. Using available data for $S$ vs. $\sigma$ for n-$Bi_2Te_3$ and p-$Bi_2Te_3$ [29], we perform full numerical simulations assuming DOS scattering and adjust the electron-phonon coupling term in (4) to match the bipolar experimental data. As shown in the Supplementary Information, an excellent fit is obtained if a bandgap of $E_G = 0.18$ eV is assumed. Figure 8, shows the resulting $L$ vs. $S$ characteristic for $Bi_2Te_3$ at 300 K calculated with and without bipolar effects included.

To understand the shape of the bipolar curve in Fig. 8, consider the movement of Fermi level beginning from inside valence band and moving toward the middle of the band gap. Deep inside the valence bands, bipolar effects due to conduction band electrons are negligible. In this case, $L$ approaches the degenerate limit of $\pi^2/3$, and $S$ is small. This is the left part of the plot where $S$ is small and both bipolar and unipolar $L$ vs. $S$ characteristics converge. As the Fermi level moves above the valence maximum and toward midgap, $S$ increases and $L$ decreases, until bipolar effects become significant, $L$ begins to increase, and the bipolar characteristic diverges from the unipolar characteristic. For $S \approx 200 \, \mu V/K$, we would deduce $L' \approx 1.75$ from the unipolar characteristics and $L' \approx 2.1$ from the bipolar curve. Unless carefully modeled, bipolar effects will introduce a significant uncertainty in the estimation of the Lorenz number from a measured Seebeck coefficient. An accurately computed $L$ vs. $S$. characteristic could provide a good way to deduce the Lorenz number from the measured Seebeck coefficient, but the mapping of $L$ to $S$ is greatly complicated when bipolar effects are present. Accurate deduction of the Lorenz number would require careful modeling of bipolar effects or their suppression.

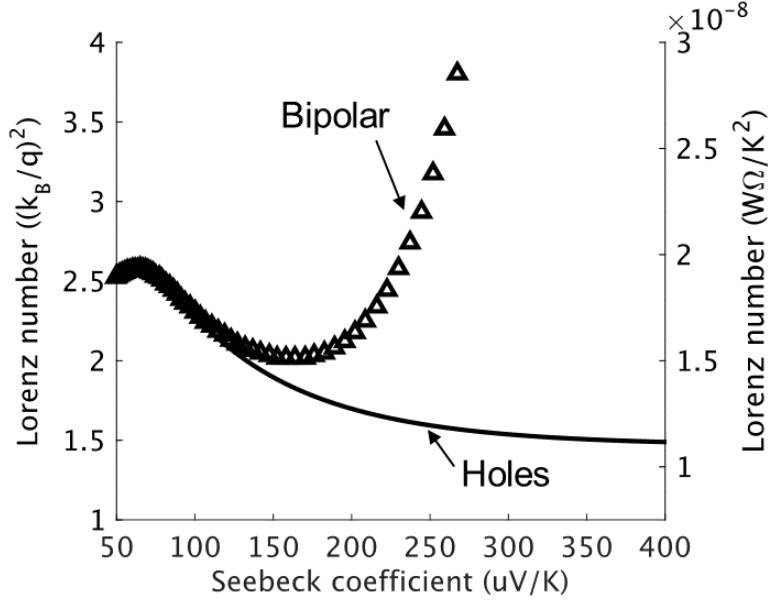

Fig. 8 The *L* vs. *S* characteristic for Bi$_2$Te$_3$ with and without bipolar conduction included ("Bipolar", symbols) and without bipolar conduction ("Holes", solid line). DOS scattering is assumed in both cases and the electron-phonon coupling constants are evaluated as discussed in the Supplementary Information.

*Numerical issues*

As illustrated by the examples presented, accurate calculations of *L* vs. *S* require a careful treatment of band structure and the energy-dependent scattering time. Numerical issues are also important. A good resolution of band structure and scattering rates near the band edges is essential. Even with a highly refined *k*-mesh, however, obtaining accurate results for quantities near the band edge can still be challenging. This can be seen in Fig. 2c, the MFP calculated numerically for silicon. The MFP shown in Fig. 2c is very close to the band edge where the scattering rate, which is proportional to the DOS, rapidly approaches zero. The MFP is proportional to velocity times the scattering time (the inverse of the scattering rate). Near the band edge, the velocity approaches zero and the scattering time approaches infinity, which presents numerical challenges. The results displayed in Fig. 2c used a refined *k*-mesh produced with an iterative technique that is described in the Supplementary Information. The numerical fluctuations in the MFP path are however, rather small, and what matters finally is the transport function, which is the product of MFP and number of conducting channels. As shown in Fig. 2d, the transport function is smooth near the band edge. Finally, we note that when an exactly parabolic band structure is used in the numerical procedure, the results are in close agreement with the analytical results, which indicates that the iterative *k*-grid refinement procedure described in the Supplementary Material is effective.

## 5. Summary

In this paper, we presented a technique to compute the Lorenz number for complex thermoelectric materials. The method makes use of DFT-calculated band structures and assumes acoustic deformation potential scattering (or more generally, a scattering rate that follows the density-of-states). The accuracy of the numerical method was established by comparing the computed *L* vs. *S* characteristic for a material (Si) with nearly parabolic energy bands to analytical results that assume parabolic bands and a constant mean-free-path. We also showed that when the transport distribution increases less slowly than linearly with energy away from the band edge, then non-degenerate Lorenz numbers less than $2(k_B/q)^2$ result.

All thermoelectric transport coefficients are determined by the transport function [10]; the Lorenz number is determined by the shape, but not the magnitude of the transport distribution. It is sensitive, therefore, to both the band structure and to energy dependent scattering. It is easy to show mathematically from eqn. (10b) that a delta-function transport distribution gives $L=0$. The physical reason is clear. If there is a single channel, then when we open-circuit it to measure $\kappa_e$, no electrons flow. Since there is no flow of electrons, there can be no flow of heat. If a transport distribution decreases rapidly with energy away from the band edge, then the transport distribution approximates a delta-function at the band edge, and low Lorenz numbers can be expected. More generally, we showed that whenever the transport function increases less than linearly with energy, a non-degenerate Lorenz number less than the parabolic band with ADP scattering result of $L'=2$ can be expected. Note also that the influence of energetically offset bands on *L* as discussed by Thesberg, et al. [23] can also be understood in terms of the transport distribution.

It has been shown that when parabolic bands with acoustic deformation potential scattering is a good approximation, then the *L* vs. *S* characteristic provides a convenient way to deduce the Lorenz number from a measured Seebeck coefficient [7]. As illustrated in this paper, the *L* vs. *S* characteristic is material-specific. We presented a procedure to compute material-specific Lorenz numbers and *L* vs. *S*. characteristics using the best available band structures and scattering rates. We illustrated this general procedure with some specific examples.

The ability to compute a material-specific *L* vs. *S* characteristic with high confidence would be useful in two ways. First, it could be used to deduce *L* from the more readily measured *S*, which would be useful for analyzing measurements of total thermal conductivity in order to deduce the lattice component. Second it could be used to identify materials that give a low *L* for a given *S*. Such materials might be useful for increasing the thermoelectric figure of merit, *zT* [1]. To produce such high confidence calculations requires high accuracy band structures (especially near the band edges), and high-accuracy energy-dependent scattering times and mean-free-paths. Decades of work on band structure calculations suggests that the first challenge can be solved, but computing the energy-dependent mean-free-path for arbitrary, complex thermoelectric materials is still not routine. It will be important to test the assumption that acoustic deformation potential scattering is generally the dominant scattering mechanism in thermoelectric materials [24]. First-principles scattering calculations may provide the capabilities needed [10-14, 22].

*Acknowledgement* – This work was partially supported by the Defense Advanced Research Projects Agency (Award No. HR0011-15-2-0037). JM acknowledges support from NSERC (Discovery Grant RGPIN-2016-04881) and Compute Canada.